\documentclass[aps, prb, groupaddress, reprint, showpacs]{revtex4-1}

\usepackage{amsfonts,tikz}
\usepackage{amsmath}
\usepackage{amssymb, bm}
\usepackage{graphicx, epstopdf, color,tikz}
\usepackage{soul}

\begin{document}

\title{Relativistic space-charge-limited current for massive Dirac fermions}

\author{Y. S. Ang}
\email{yeesin\_ang@sutd.edu.sg (Corresponding author)}

\author{M. Zubair}

\author{L. K. Ang}
\email{ricky\_ang@sutd.edu.sg (Corresponding author)}

\affiliation{SUTD-MIT International Design Center, Singapore University of Technology and Design, Singapore 487372}

\begin{abstract}

A theory of relativistic space-charge-limited current (SCLC) is formulated to determine the SCLC scaling, $J\propto V^{\alpha}/L^{\beta}$, for a finite bandgap Dirac material of length $L$ biased under a voltage $V$. 
In a one-dimensional (1D) bulk geometry, our model allows ($\alpha$, $\beta$) to vary from (2,3) for the non-relativistic model in traditional solids to (3/2,2) for the ultra-relativistic model of massless Dirac fermions.
For a two-dimensional (2D) thin-film geometry, we obtain $\alpha = \beta$ that varies between 2 and 3/2, respectively, at the non-relativistic and ultra-relativistic limits.
We further provide a rigorous proof based on a Green's function approach that for uniform SCLC model described by carrier density-dependent mobility, the scaling relations of the 1D bulk model can be directly mapped into the case of 2D thin film for any contact geometries.
Our simplified approach provides a convenient tool to obtain the 2D thin-film SCLC scaling relations without the need of explicitly solving the complicated 2D problems.      
Finally, this work clarifies the inconsistency in using the traditional SCLC models to explain the experimental measurement of 2D Dirac semiconductor. We conclude that the voltage-scaling $3/2 < \alpha < 2$ is a distinct signature of massive Dirac fermions in Dirac semiconductor and is in agreement with experimental SCLC measurement in MoS$_2$.


\end{abstract}

\maketitle


\section{Introduction} 
Space-charge-limited current (SCLC) gives the maximum current that can be transported across a solid of length $L$ with a biased voltage $V$, limited by the electrostatic repulsion generated by the \emph{in-transit} unscreened charge carriers that are in excess of the thermodynamically allowed population \cite{lampert}. 
In a trap-free bulk crystal, SCLC exhibits a signature current-voltage ($J$-$V$) characteristics of $J_{MG}\propto V^2 / L^{3}$ known as the Mott-Gurney (MG) law \cite{MG}, which is the solid-state counterpart of the SCLC in vacuum as given by the Child-Langumir (CL) law: $J_{CL}\propto V^{3/2} / L^{2}$ in classical regime \cite{CL,Zhu} and $J_{CL}\propto V^{1/2} / L^{4}$ in quantum regime \cite{Ang, Lau}. 
Including defect states or traps in solids, SCLC becomes \emph{trap-limited} as described by the Mark-Helfrich (MH) law \cite{mark}: $J_{MH} \propto V^{l+1}/L^{2l+1}$, where $l = T_c/T$, $T$ is temperature and $T_c$ is a parameter characterizing the exponential spread in energy of the traps. 
Due to the geometrical effect \cite{grinberg}, the 1D SCLC value is enhanced as a result of finite emission area \cite{chandra} and weakened Coulomb screening in high aspect-ratio nanowire \cite{talin}. 
Furthermore, SCLC is an important tool to probe the trap characteristics in solids, and also for photocurrent measurement since the extraction efficiency of photogenerated carriers is fundamentally limited by SCLC \cite{SCLPC}.

For organic semiconductors, field-dependent \cite{blom, bozano, campbell, ng} and density-dependent mobility SCLC models \cite{tanase, pasveer} are commonly employed to characterize the SCLC carried by the holes.
Similarly, SCLC of electrons was found to be universally described by a trap model with Gaussian energy distribution in a large class of organic semiconductors \cite{nicolai, nicolai2}. 
Recently it is demonstrated that the magnitude of electron SCLC can be significantly enhanced via the dilution of traps in conjugated polymer blends of only 10\% of active semiconductor \cite{abbaszadeh}, which opens up an exciting possibility of high-efficiency and low-cost organic light emitting diode. 
SCLC in the trap-limited regime was re-formulated \cite{zhang} with inclusion of the interplay between dopants and traps, Poole-Frenkel effect \cite{frenkel} and quantum mechanical tunneling, which has solved the long-standing problem \cite{rose, lampert2} of the enormously sharp current rise at the trap-limited regime and demonstrated that an exponentially distributed trap is not necessarily required to explain the power-law sharp rises of SCLC in the trap-limited regime. Remarkably, the model successfully reproduced the anomalous noise-spectrum peak observed in \cite{carbone}.

In spite of SCLC being a classic model first derived in 1940s, it remains an active topic for organic materials and nanowires as mentioned above. 
With the advances in fabricating novel 2D Dirac materials \cite{graphene, TI, mos2_lit}, it is of the interests to revisit the SCLC model for these 2D Dirac materials. 
To our best knowledge, there is no theory or model to deal with the SCLC transport in Dirac materials.
Recent experiments reported a typical $J$-$V$ characteristic in the form of MH law for highly disordered materials like reduced graphene oxide \cite{jung, jung2}.
On the other hand, SCLC in crystalline monolayer MoS$_2$ \cite{mos2} and hBN \cite{hbn} was found to exhibit an unusual power law dependence of $J \propto V^\alpha$ with $1.7\lesssim\alpha\lesssim2.5$, which was claimed to be originated from the different levels of traps in different samples by using the traditional MH law.
This explanation is doubtful as the traditional MH law is only valid for $\alpha>2$ for $T_c>T$, which implies that the voltage scaling from the MH law must be $\alpha \geq 2$ theoretically and it can not be used to fit with the mesured scaling of $1.7\lesssim\alpha\lesssim2.5$.
For $T_c<T$, the traps are narrowly distributed in energy space and the SCLC essentially reduces to single-level shallow trapping with $\alpha = 2$ \cite{rose}. 
Thus, the observation of $\alpha<2$ can not be explained by the MH law or other SCLC models such as shallow trap \cite{rose}, Gaussian disorder \cite{gdm}, field-dependent \cite{blom} and density-dependent \cite{tanase} mobility. 

For a Dirac material with finite bandgap, the electrons mimics relativistic \emph{massive Dirac fermions} \cite{mos2_band, hbn_band} whereas the classical SCLC models are based on the conduction model of \emph{non-relativistic} quasi-free electrons \cite{lampert}. 
In this work, we proposed a model of \emph{relativistic} SCLC of massive Dirac fermions, which can explain the peculiar $\alpha<2$ scaling observed in recent experiments using Dirac materials and thus circumvents the un-justified assumption of $T_c<T$ used in the MH law in order to fit the experimental data. 
According to our model, the $J_1$-$V$ characteristics of SCLC in a 1D bulk geometry will vary between the non-relativistic limit of $J_1\propto V^2/L^3$ to the ultra-relativistic limit of $J_1\propto V^{3/2}/L^2$ (this is different from the CL law - see below for explanation).
We present a master equation which is in good agreement with the experimental data and can be used to characterize the transition between the Ohmic conduction and SCLC regime.
By extending the bulk 1D model to a 2D thin film model, the scaling relation becomes $J\propto V^{\alpha}/L^{\beta}$ with $\alpha=\beta$ varying between 3/2 and 2, respectively, at the ultra-relativistic and non-relativistic limits.
In doing so, we prove rigorously, using a Green's function approach\cite{grinberg}, that the 1D bulk SCLC current-voltage scaling relation can be directly mapped to 2D thin-film SCLC. 
It is shown that for a general transport equation of $J=en\mu(n)E$ where $\mu(n)$ is a mobility that depends on carrier density, $n$, and $E$ is the electric field, the 1D bulk SCLC and 2D thin-film SCLC are linked by a universal SCLC scaling relation (see Section III).
Our analysis provides a convenient tool to deduce the 2D thin-film SCLC scaling relation via simple 1D SCLC model without the need of explicitly solving the complicated 2D SCLC model.


\section{Theory of relativistic space-charge-limited current}

In this section, a relativistic SCLC model is developed using semiclassical Boltzmann transport equation (BTE). For simplicity, we first consider the SCLC by assuming a simple 1D Poisson equation, which allows semi-analytical scaling relations to be determined. In Section III, we shall show that the simple 1D SCLC scaling relation derived in this section can be directly mapped to the case of 2D SCLC with thin-film geometry.
\subsection{Boltzmann transport equation for conventional semiconductor}

The starting point of the trap-free SCLC theory, i.e. the Mott-Gurney's law, is the semiclassical BTE which provides a basic equation of current density governing the transport of charge carriers. The diffusion component is usually not considered except in some cases of polymers due to their highly disordered nature. For a quasi-static system, the BTE in the linearized transport regime under relaxation-time approximation is
\begin{equation}
-\frac{e\mathbf{E}}{\hbar} \cdot \frac{\partial f}{\partial \mathbf{k} } + \mathbf{v} \cdot \frac{\partial f}{\partial \mathbf{r}} = - \frac{f - f_0}{\tau}
\end{equation}
where $\mathbf{E}$ is the electric field, $\mathbf{k}$ is the crystal momentum, $\mathbf{r}$ is the position vector, $\mathbf{v}$ is the carrier velocity, f is the out-of-equilibrium distribution function, $f_0$ is the equilibrium Fermi-Dirac distribution function, and $\tau$ is a typical collision time scale.
If the system is spatially homogeneous the diffusion component of the transport current, i.e. $\partial f / \partial \mathbf{r}$, can be omitted. By assuming a 3D isotropic parabolic energy dispersion, one arrives at the well-known drift-current density, $J=e/(3\pi)^3 \int v_k f d^3  \mathbf{k}  $, for semiconductors:
\begin{equation}
J_{3D} = \frac{\tau e^2 }{m}n(x) E(x)
\end{equation}
By connecting the drift-equation with the 1D Poisson equation via charge density $n(x)$, the Mott-Gurney current-voltage scaling of $J\propto V^2$ can be recovered.

For 2D gapped Dirac materials, Eq. (2) is no longer valid due to two reasons: (i) the dimensionality is reduced to 2D; and (ii) the energy dispersion follows a relativistic dispersion similar to that of the massive Dirac fermions. In the following, we shall formulate the drift-equation for Dirac semiconductor based on BTE approach and demonstrate that the SCLC mediated by relativistic quasiparticles follows a completely different current-voltage scaling relation.

\subsection{Boltzmann transport equation for 2D Dirac semiconductor}

For massive Dirac fermions, the energy dispersion is $\varepsilon_k = \sqrt{\hbar^2v_F^2k^2 +\Delta^2}$ where $v_F$ is the Fermi velocity, $k$ is the crystal momentum and $2\Delta$ is the bandgap.  
The group velocity is $v_k = \hbar^{-1} d \varepsilon_k /dk = \hbar v_F^2k/\varepsilon_k$. 
The density of states $D(\varepsilon) = \sum_{\mathbf{k}} \delta(\varepsilon -\varepsilon_k)$, is rewritten as $D(\varepsilon) = (g_{sv} \varepsilon/2\pi\hbar^2v_F^2) \Theta(\varepsilon - \Delta)$ where $g_{sv}$ denotes the spin-valley degeneracy and $\Theta(x)$ is a Heaviside function. The electron density at low temperature can then be obtained from the two-dimensional (2D) density of states $n=\int D(\varepsilon_k) d\varepsilon_k$ which gives $	n = (g_{sv}/4\pi\hbar^2v_F^2)  \left( \mu^2 - \Delta^2 \right)$ and $\mu$ is the Fermi level. 
The general expression of the 2D linear current density is $J = (\tau e^2E/2\pi) \int v_k^2 kdk \left(-\partial f_0/ \partial \varepsilon_k\right)$ where $f_0$ is the Fermi-Dirac distribution function, $\tau$ is the scattering time and $E$ is the externally applied electric field. 
In the low temperature limit, the current density can be analytically solved to give $J = (g_{sv}\tau e^2E/2\pi\hbar^2\mu) (\mu^2 - \Delta^2)$. 
Eliminating $\mu$ via $n$, we obtain
\begin{equation}\label{ce}
J = \sqrt{\frac{eg_{sv}}{\pi}} \frac{\tau e v_F}{\hbar} \frac{en(x)}{\sqrt{en(x)  +  \rho_c }} E(x),
\end{equation}
where $\rho_c \equiv eg_{sv}\Delta^2/4\pi\hbar^2v_F^2 $ is a bandgap-dependent \emph{characteristic charge density}, $n$ and $E$ are re-expressed as functions of the transport direction, $x$. The term $\left(en(x) + \rho_c\right)^{-1/2}$ in Eq. (\ref{ce}) represents a major difference between the relativistic massive Dirac fermions and that of the non-relativistic quasi-free electrons. 
As $\rho_c \propto \Delta^2$, it can be seen from the $\varepsilon_k$-$k$ relation that the electrons are non-relativistic at very large $\rho_c$ or $\Delta \gg \hbar v_F k$. 
For vanishingly small $\rho_c$ or $\Delta \ll \hbar v_Fk$, the electrons approach ultra-relativsitic limit and become \emph{massless} Dirac fermions.

By expressing Eq. (1) in the Drude form of $J = en\mu_{D}(n)E$, a density-dependent \emph{Dirac mobility} is defined as $\mu_D \equiv \gamma \left( en(x) + \rho_c \right)^{-1/2}$ where $\gamma \equiv \tau v_F e^{3/2} g_{s,v}^{1/2}/\pi^{1/2}\hbar$. 
Consequently, the relativistic SCLC of massive Dirac fermions belongs to the class of \emph{density-dependent mobility} SCLC.
However, $\mu_D \equiv \gamma \left( en(x) + \rho_c \right)^{-1/2}$ is unique to the massive Dirac fermions.

\subsection{Relativistic SCLC in bulk geometry}

We assume that the SCLC is carried solely by electrons injected through an Ohmic contact. 
For simplicity, we employ the 1D Poisson equation $dE(x)/dx = en(x)/\epsilon d$ where $E(x) = dV(x)/dx$, $V(x)$ is the electrostatic potential, and $\epsilon$ is the effective dielectric constant. Here, we first assume that the 3D carrier density, $n_{3D}(x)$, is related to $n(x)$ via $n_{3D}(x) = n(x) / d$ with $d$ as an effective thickness (see modification later).
By coupling Eq. (\ref{ce}) with the Poisson equation via $n(x)$, we obtain the governing equation of the relativistic SCLC for Dirac solid:
\begin{equation}\label{sclc}
E(x) \frac{dE(x)}{dx} = \frac{J}{\gamma \epsilon d} \sqrt{ \epsilon d \frac{dE(x)}{dx}  + \rho_c}.
\end{equation}

We first investigate the solutions of the 1D relativistic SCLC, i.e. Eq. (\ref{sclc}), in two asymptotic limits: (i) non-relativistic SCLC regime ($\rho_c \gg en(x)$); and (ii) ultra-relativistic SCLC regime ($\rho_c \ll en(x)$),
which allows Eq. (\ref{ce}) to be approximated, respectively, as
\begin{subequations}\label{lim}
\begin{align}
J_{nr} &= \frac{9}{8} \epsilon d \frac{2 \tau e v_F^2}{\Delta} \frac{V^2}{L^3} & \text{, }\rho_c \gg en(x) \label{lim1}, \\
J_{r} &= \frac{8}{3}\sqrt{\frac{eg_{sv}\delta\epsilon}{3\pi}} \frac{\tau e v_F}{\hbar} \frac{V^{3/2}}{L^2} & \text{, } \rho_c \ll en(x) \label{lim2}.
\end{align}
\end{subequations}
The MG scaling is readily recovered from the non-relativistic charge dynamics at large $\rho_c$. 
As the electrons reside just slightly above the bandgap where the inequality $\hbar v_Fk \ll \Delta$ holds true, we have $\varepsilon_k \approx m^* v_F^2 + \hbar^2 k^2/2 m^*$ where $m^* \equiv \Delta/v_F^2$. 
It can be shown \cite{fn1} that the corresponding current density is in the non-relativistic Drude form, which recovers the MG scaling of $J\propto V^{2}/L^3$. 

In the opposite limit of $\rho_c \to 0$, $\hbar v_F k \gg \Delta$ implies ultra-relativistic dynamics with a scaling of $J\propto V^{3/2}/L^2$ as shown in Eq. 3(b).
Coincidentally, this has the same scaling to the CL law \cite{CL} although the underlying physics is fundamentally different. 
For Dirac materials studied there, the ultra-relativistic SCLC is obtained from
\begin{equation}\label{dense}
	J_{r} \propto \sqrt{ \frac{d^2V(x)}{dx^2} } \frac{dV(x)}{dx},
\end{equation}
while SCLC in vacuum (or CL law) is obtained from
\begin{equation}\label{cl}
	J_{vac} \propto \sqrt{ V(x)} \frac{d^2V(x)}{dx^2} .
\end{equation}
A simple dimensional analysis \cite{ang2} immediately shows that $J_{r}$ and $J_{vac}$ are both proportional to $V^{3/2}$ albeit their very different origins. 
Nonetheless, Eq. (\ref{dense}) originates from the $J_{r} \propto \sqrt{n(x)}$ dependence owing to the \emph{ultra-relativistic} electron dynamics in Dirac solids while Eq. (\ref{cl}) originates from the $J_{vac} \propto \sqrt{V(x)}$ dependence owing to the energy balance of a \emph{non-relativistic} free electron accelerating in vacuum. 
This demonstrates the fundamentally different mechanism behind the $J\propto V^{3/2}$ scaling in the two cases. 

The two limits: $\alpha = 3/2$ and $\alpha = 2$ are, respectively, the extreme limits of the ultra-relativsitic and non-relativistic SCLC, an intermediate regime of $3/2<\alpha<2$ is expected in the case of massive Dirac fermions. 
This is in good agreement with the experimental observations of $1.7<\alpha<2.5$ in monolayer MoS$_2$ \cite{mos2} and $1.75<\alpha<2.5$ in monolayer hBN \cite{hbn} where the charge carriers are essentially massive Dirac fermions. 
The model proposed here suggests that the $\alpha<2$ scaling is $intrinsic$ to the relativistic carriers without the unjustified or invalid assumption of introducing traps with $T_c<T$, which is also inconsistent with the original formulation of the MH law and other trap-limited SCLC models as discussed above.

\begin{figure}[t]
	\includegraphics[scale=0.65]{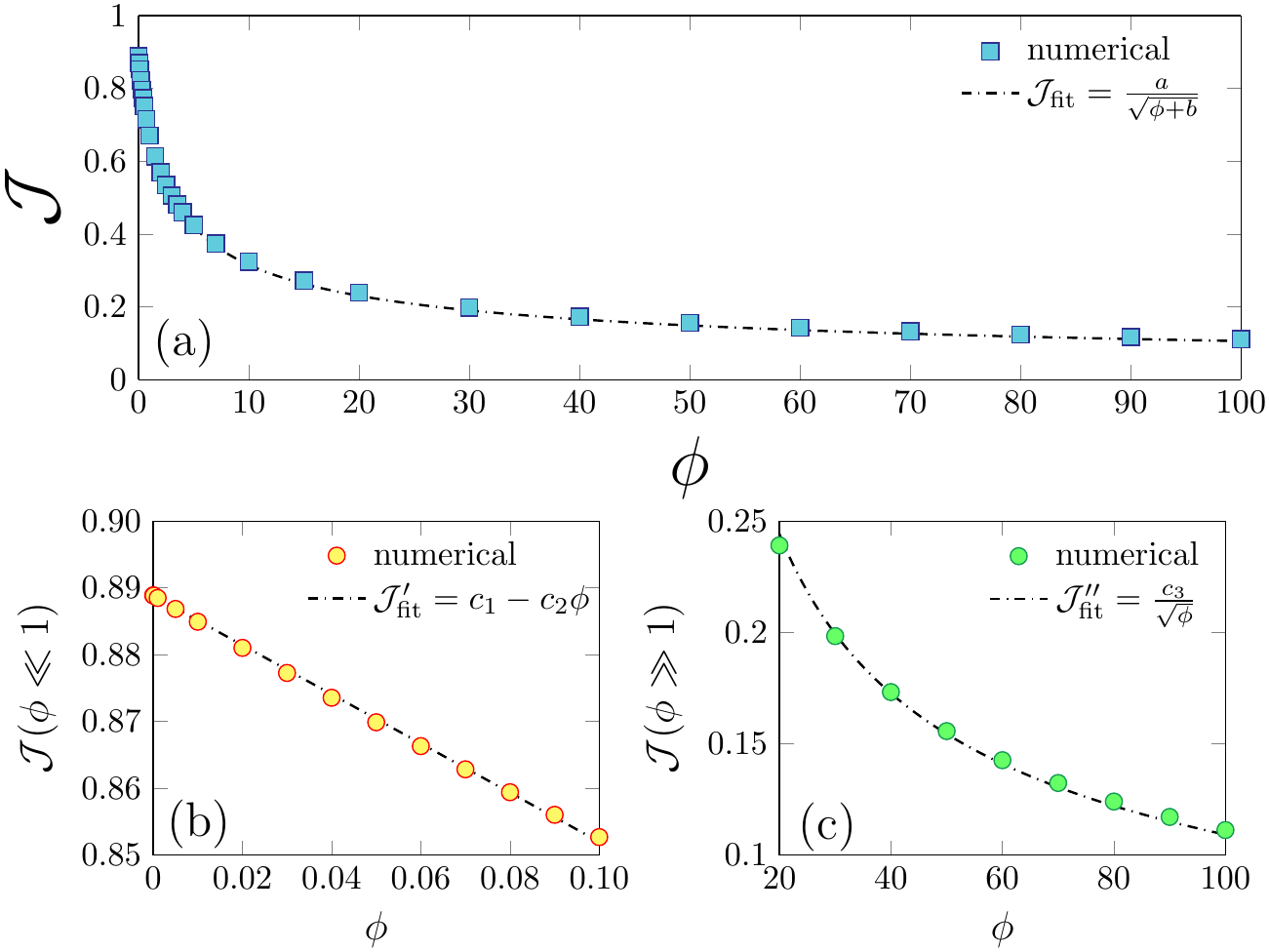}
	\caption{$\mathcal{J}$ as a function of $\phi$. (a) Numerical results of $\mathcal{J}$ over the full range of $\phi$. The dashed line shows the empirical fitting equation; numerical results with (b) $\phi \ll 1$; and at (c) $\phi \gg 1$. The fitting constants ($a$, $b$, $c_1$, $c_2$, $c_3$) are (1.067, 1.45, 0.889,0.368,1.092).
	}
\end{figure}

For convenience, we transform Eq. (\ref{sclc}) into a dimensionless form of
\begin{equation}\label{sclc2}
	\frac{d \mathcal{V}(\chi) }{d \chi} \frac{d^2 \mathcal{V}(\chi) }{d \chi^2} = \mathcal{J} \sqrt{\frac{d^2 \mathcal{V}(\chi) }{d \chi^2}  + \phi },
\end{equation}
where $\mathcal{V}(x) \equiv V(x)/V$ and $\chi \equiv x/L$. 
The normalized current $\mathcal{J}$ and the dimensionless parameter $\phi$ are 
\begin{subequations}\label{para}
	\begin{align}
		\mathcal{J} &\equiv \frac{\hbar}{\tau e^{3/2}v_F \sqrt{\epsilon d} } \frac{JL^2}{V^{3/2}}, \\
		\phi &\equiv \frac{\rho_c}{\epsilon d } \frac{L^2}{V}.
	\end{align}
\end{subequations}
For a given value of $\phi$ and boundary conditions: $\mathcal{V}(0) = 0$ and $\mathcal{V}(1) = 1$, Eq. (\ref{sclc2}) is solved numerically at various $\mathcal{J}$.
The corresponding space charge limited current (SCLC) is determined when the value of $\mathcal{J}$ will cause the onset of $\mathcal{V}(\chi)<0$, and the calculated SCLC $\mathcal{J}$ is plotted as a function of $\phi$ in Fig. 1(a), which exhibits contrasting behaviour at $\phi\ll 1$ and at $\phi \gg 1$.
For $\phi \ll 1$ [Fig. 1(b)] and $\phi \gg 1$ [Fig. 1(c)], the numerical results can be fitted by $\mathcal{J}_{\text{fit}}' = c_1 - c_2 \phi$ and $\mathcal{J}_{\text{fit}}'' = c_3/\sqrt{\phi}$, respectively, where $(c_1, c_2, c_3) = (0.889, 0.368, 1.092)$. 
The contrasting $\phi$-dependence can be understood from the dependence of $\phi \propto \rho_c$, which corresponds, respectively, to the ultra-relativistic SCLC 
at  $\phi \ll 1$ (or small $\phi$), and the non-relativistic SCLC at $\phi \gg 1$ (or large $\phi$). 
By substituting Eq. (\ref{para}) into the above mentioned fitting equations, $J_r\propto V^{3/2}/L^2$ and $J_{nr}\propto V^2/L^3$ are recovered, thus confirming the analytical solutions given in Eq. (\ref{lim}).

Figure 2 shows the smooth transition of $V$ and $L$ in between the ultra-relativisitc and non-relativistic regimes. 
The dimensionless current density $\tilde{J}_L$-$\tilde{V}$ characteristic (at a fixed $L$) exhibits a voltage scaling of $\alpha$ = 2 at low-$\tilde{V} $ and $\alpha$ = 3/2 at high-$\tilde{V}$.
Here the dimensionless parameters are $\tilde{J}_L \equiv \mathcal{J}/J_0$, $\tilde{V} \equiv V/V_0 $, $J_0 \equiv \tau e^{3/2}v_F\sqrt{\epsilon \delta}/\hbar$ and $V_0=\rho_cL^2/\epsilon\delta$.
At a fixed $L$, we have $\tilde{V}\propto 1/V_0 \propto 1/\rho_c$. and thus low-$\tilde{V}$ and high-$\tilde{V}$ correspond to the non-relativistic ($\tilde{J}_L \propto \tilde{V}^2$) and the ultra-relativistic ($\tilde{J}_L \propto \tilde{V}^{3/2}$) regimes, respectively as shown in Fig. 2 (blue axis, $\square$-symbols). 
In the intermediate regime, $\tilde{J}_L$-$\tilde{V}$ deviates from the simple power law, and applying a fitting would lead to a sub-quadratic scaling in the range of $3/2 < \alpha < 2$. 

Similarly, the dimensionless $\tilde{J}_V$-$\tilde{L}$ characteristic (at a fixed $V$) shows a length scaling of  $J\propto \tilde{L}^{-\beta}$ of $\beta = 2$ at small $\tilde{L}$ and $\beta = 3$ at large $\tilde{L}$, as shown in Fig. 2 (green axis, $\triangle$-symbols).
Here, the dimensionless parameters are $\tilde{J}_V = J/\bar{J}_0$, $\tilde{L} = L/L_0$, $\bar{J}_0 = \tau e^{3/2}v_F \rho_c V^{1/2}/ \hbar \sqrt{\epsilon d}$ and $L_0  = \sqrt{\epsilon d V/\rho_c}$. 
As $\tilde{L} \propto \sqrt{\rho_c}$, small $\tilde{L}$ and large $\tilde{L}$ corresponds to the non-relativistic and the ultra-relativistic regimes, respectively. 
The $\beta<3$ sub-cubic inverse length scaling represents another signature of the relativistic SCLC for Dirac solids in addition to the $\alpha<2$ voltage scaling.

\begin{figure}[t]
	\includegraphics[scale=1]{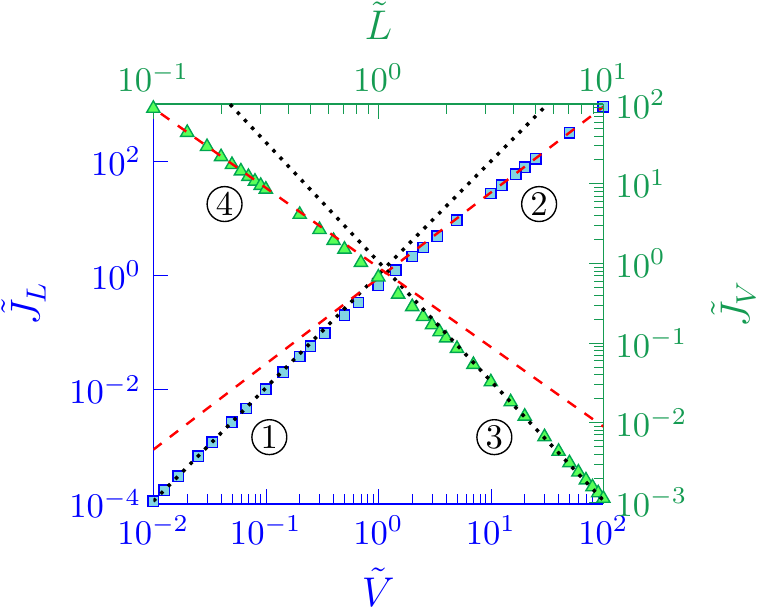}
	\caption{Normalized voltage (blue axes, $\square$-symbols) and length (green axes, $\triangle$-symbols) characteristics of the current density. The broken (red) and the dotted (black) guide-line represents ultra-relativistic and non-relativistic limit, respectively. The labels of the guide-lines, i.e. 1, 2, 3 and 4, correspond to the scalings $\tilde{J}_L\propto V^{2}$, $\tilde{J}_L\propto V^{3/2}$, $\tilde{J}_V \propto L^2$ and $\tilde{J}_V \propto L^3$, respectively. }
\end{figure}

From Fig. 1(a), the numerical results over a wide range of $\phi$ can be accurately fitted by $\mathcal{J}_{\text{fit}}(\phi) = \frac{a}{\sqrt{\phi+b}}$ where $(a,b) = (1.067, 1.450)$. 
As all parameters are intrinsically contained in $\mathcal{J}$ and $\phi$, this empirical relation is universally valid and thus we derive a \emph{master equation} that universally describes the SCLC transport over a wide range of parameters:
\begin{equation}\label{emp3}
\frac{V^3}{I^2} = \frac{\Lambda(\rho_c,\epsilon d,\tau, L, W)}{V} + \Omega(\epsilon d, \tau, L, W),
\end{equation}
where $I = J \times W$ is the total current, $W$ is the device width, and 
\begin{subequations}
	\begin{align}
	\Lambda(\rho_c,\epsilon d,\tau, L, W) &\equiv \frac{\rho_c}{\left(\epsilon d\right)^{2}} \frac{\hbar^2L^{6}}{a^2\tau^2v_F^2e^3 W^2}, \\
	\Omega(\epsilon d, \tau, L, W) &\equiv \frac{b\hbar^2L^4}{a^2\tau^2v_F^2e^3 W^2} \frac{1}{\epsilon d}.
	\end{align}
\end{subequations}
It is important to emphasize that Eq. (\ref{emp3}) is extremely usefully if it is used to fit with the experimental $I$-$V$ measurement (in the form of $V^3/I^2$ as a function of $1/V$) to determine the values of $\Lambda$ and $\Omega$, which can be subsequently used to determine the collision time scale $\tau$ by using Eqs. (9) if the other parameters are known.

For the ultra-relativistic limit at $\rho_c\to0$, Eq. (\ref{emp3}) becomes $V^3/I^2 \approx \Omega$ which confirms the $predicted$ ultra-relativistic scaling of $(\alpha,\beta) = (3/2,2)$. 
For the non-relativistic limit at $\rho_c \gg 0$, Eq. (\ref{emp3}) reduces to $V^3/I^2 \approx \Lambda/ V$ which recovers the classical MG scaling of $(\alpha,\beta) = (2,3)$ as expected.
Therefore, the intermediate relativistic SCLC will produce a positive intercept on the vertical-axis of $V^3/J^2$-$1/V$ characteristic whereas the SCLC with non-relativistic scaling will have a zero intercept as shown in Fig. 3(b).

Interestingly, the $V^3/I^2$-$1/V$ characteristic [suggested in Eq. (10)] provides a convenient tool to represent the SCLC data that can be generally applied to any solids. 
To illustrate this point, we consider a trap-free solid in which the conduction transits from Ohmic to SCLC at increasing $V$ as showin in Fig. 3a.
In the Ohmic regime (large $1/V$ or small $V$) where $J\propto V$, we have $V^3/I^2 \propto (1/V)^{-1}$, i.e. $V^3/I^2$ decreases with increasing $1/V$ (green dash-dotted line in Fig. 3a).
In contrast, in the SCLC regime (small $1/V$ or large $V$) where $I\propto V^2$, we have $V^3/I^2 \propto 1/V$, i.e. $V^3/I^2$ increases linearly with $1/V$ (blue dashed line in Fig 3a). 
These contrasting behaviors lead to a $transitional$ peak \cite{fn2} in the intermediate regime that clearly separates the SCLC-dominated and Ohmic-dominated conduction as shown in Fig. 3(a).
This finding is confirmed by using various experimental data (color symbols) for MoS$_2$ from Ref. \cite{mos2} at different temperature as shown in Fig. 3a.
The above mentioned transitional peak between SCLC-dominated regime (dashed curve) and Ohmic-dominated regime (dash-dotted curve) can be clearly observed at all temperatures. 

A zoom-in view at the small-$1/V$ SCLC-dominated regime is shown in Fig. 3(b), which indicates that the experimental results (symbols) can be explained by linear fitting so to obtain the voltage scaling, which ranges from $\alpha$ = 2.11 down to 1.67 according to Eq. (\ref{emp3}).
The $\alpha$ scaling decreases to $\alpha<2$ at elevated temperature because the higher energy levels of the conduction band is increasingly populated by the thermally-liberated electrons from trapping sites.
This leads to a higher-degree of relativistic dynamics of the transport electrons thus reducing $\alpha$. 
For the three $\alpha<2$ cases, they can be extrapolated to have positive intercepts on the $y$-axis at $1/V \to$ 0 [see Fig. 3(b)].
As $\alpha$ approaches 2, the intercepts diminishes and becomes approximately zero at $\alpha = 2$. 
These observations are in good agreement with the predicted intercepts of Eq. (\ref{emp3}), as discussed above. 
Note that Eq. (\ref{emp3}) breaks down in the case of $\alpha>2$ where the intercept becomes negative. 

\begin{figure}[t]
	\includegraphics[scale=1]{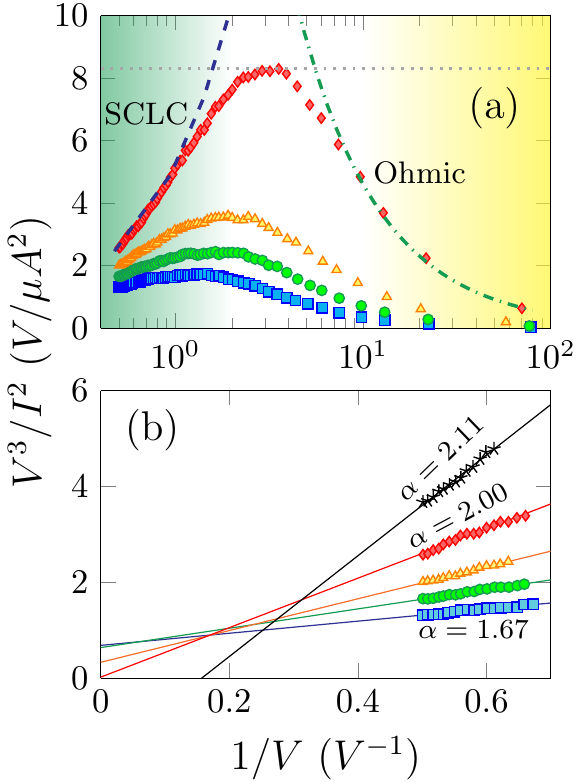}
	\caption{Plot of $V^3/I^2$ against $1/V$ in the SCLC regime using MoS$_2$ experimental data from Ref. \cite{mos2} with (T,$\alpha$) of (285K, 1.67) (blue square), (265K, 1.73) (green circle), (245K, 1.82) (yellow triangle), (205K, 2.00) (red diamond) and (185K, 2.11) (black star). (a) The entire $V^3/I^2$ range over 0.4 $<1/V<$ 100. The green dash-dotted and blue dashed curves denote, respectively, Ohmic current ($J\propto V$) and SCLC ($J\propto V^2$) fitted to the $T= 205$ K data. The horizontal gray line indicate the transitional regime of $J\propto V^{3/2}$ (1 $<1/V <$ 6) separating the green-shaded SCLC-dominated and the yellow-shaded Ohmic-dominated regime. (b) The SCLC-dominated regime at small $1/V<$ 0.7.} 
\end{figure}

\section{Simplified model of uniform SCLC injection in 2D Dirac semiconductor} 

The relativistic SCLC model derived above is based on solving the 1D Poisson equation. As Dirac semiconductor is a 2D thin film, thus a 2D thin-form model is required. In this Section, we provide a simplified formalism of 2D thin-film relativistic SCLC model without the need to explicitly solve for the 2D model \cite{grinberg}. 

\subsection{Universal model of $D$-dimensional uniform SCLC injection in solid with density-dependent mobility}

The SCLC has been previously formulated for thin-film and nanowire using an integral form of 2D electrostatic Poisson equation \cite{grinberg,talin}. Here, we shall formulate a thin-film SCLC relativistic model under the assumption of carrier density-dependent mobility to illustrate that the general SCLC scaling properties for Dirac semiconductor, which has a density-dependent mobility of $\mu_D = \gamma (en(x)+\rho_c)^{-1/2}$.


We consider the $D$-dimension transport in a solid with density-dependent mobility in a general form of $\mu = \mu_0 f\left[n(x)\right] / f_0$ where $f\left[n(x)\right]$ is a density-dependent term and $f_0$ is a constant factor. 
The dimensionality of $D=1$  and $D=2$ corresponds to bulk and 2D thin film, respectively. 
In the following analysis, we consider the case of \emph{uniform SCLC injection} where the two electrodes are separated by a fixed spacing of $L$ as shown in Fig. 4(a). For uniform SCLC injection along the $x$-direction, the electric field profiles, i.e. $E_D$ for bulk ($D=1$) and thin film ($D=2$) can be written, respectively, as
\begin{subequations}
	\begin{align}
	E_{1}(x) &= \frac{e}{\epsilon} \int dx' \frac{\partial G_{1}(x,x')}{\partial x}  n_{3}(x'), \\
	E_{2}(x,y) &= \frac{e}{\epsilon} \int dx' \int dy' \frac{\partial G_{2}(x,y,x', y')}{\partial x} \delta(y') n_2(x'), 
	\end{align}
\end{subequations}
where $n_{2}(x)$ and $n_3(x)$ denote the surface and volume carrier density respectively. $G_{1}(x,x')$ and $G_{2}(x,y,x',y')$ is, respectively, the 1D and 2D Green's function that are dependent on the geometry of contacts. 
Figures 4(b) and (c) shows a 2D thin film with two possible contact geometries, i.e. edge and strip contacts, respectively\cite{grinberg}. 
By eliminating the $y'$-integration via $\delta(y')$ and suppressing the argument of $y=0$ in Eq. (12b) for simplicity, Eq. (12) can be written compactly as 
\begin{equation}
	E_{D}(\xi) = \frac{e}{\epsilon} \int_0^1 d\xi' \frac{\partial G_{D}(\xi,\xi')}{\partial \xi} n_{\nu}(\xi'), 
\end{equation}
where we have introduced dimensionless variable as $\xi \equiv x/L$, and the subscript of $\nu = 2,3$ denotes surface and bulk carrier density, respectively. Consider the charge transport mechanism in Dirac solid belongs to the class of density-dependent mobility model, we assume a general current density of
\begin{equation}
	J_{D} = en_\nu(\xi) \mu_0 \frac{f\left[n_{\nu}(\xi)\right]}{f_0} E_{D}(\xi),
\end{equation}
where the subscript $D=1, 2$ denotes linear and areal current density respectively. The density-dependent mobility is given as $\mu[n_\nu(\xi)] \equiv \mu_0f[n_\nu(\xi)]/f_0$ where $f[n_\nu(\xi)]$ is a $n_\nu(\xi)$-dependent term and $f_0$ is a normalization constant. 
From Eq. (14), the bias voltage relation: $V = \int_{0}^{L} E_D(x') dx'$ is written as
\begin{equation}
	J_{D} f_0  \int_0^1 \frac{d\xi'}{n_{\nu}(\xi')f\left[n_{\nu}(\xi')\right]} = e\mu_0\frac{V}{L}.
\end{equation}
The solution of Eq. (15) gives the equation of SCLC. Its full solution require the knowledge of $n_\nu(\xi)$ over the intervals from $\xi = 0$ to $\xi = 1$, which can be obtained by solving the nonlinear integral equation in Eq. (13). Nonetheless, the scaling relations, i.e. $J_D$-$V$ and $J_D$-$L$, can be readily deduced via a simple \emph{dimensional analysis} \cite{grinberg} without explicitly solving Eqs. (13) and (15). 

\begin{figure}
	\includegraphics[scale=.4]{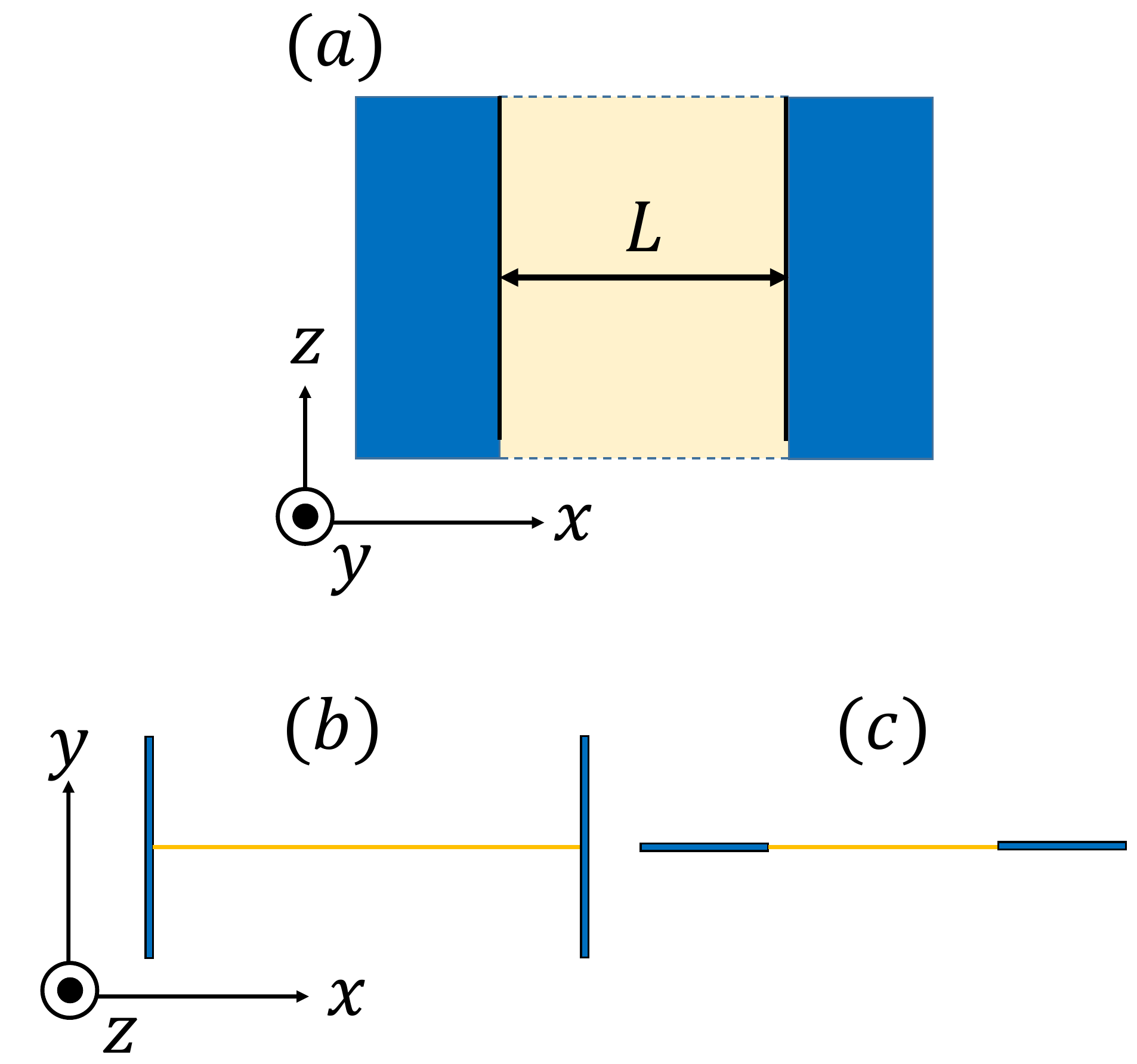}
	\caption{Schematic drawings of the device and contact geometries. Top view of (a) constant-$L$ contact geometry for uniform SCLC injection. Current is uniformly injected along the $x$-direction. For $D=1$ bulk geometry, the constant-$L$ geometry is invariant along both $z$-and $y$-directions whereas for $D=2$ thin-film geometry, the structure is only invariant along the $z$-direction. (c) and (d) shows the side view of constant-$L$ 2D thin film with edge and strip contacts, respectively.}
\end{figure}

To illustrate this, we first combine Eqs. (13) and (14) to obtain
\begin{equation}
	1 = \frac{ e^2 \mu_0 }{ \epsilon f_0 J_{D} }n_{\nu}(\xi) f\left[n_{\nu}(\xi)\right] \int_0^1 d\xi' \frac{\partial G_{D}(\xi,\xi')}{\partial \xi}n_{\nu}(\xi') .
\end{equation}
From the Poisson equation, i.e. $\nabla^2 G_D(\mathbf{r},\mathbf{r}') = \delta_D(\mathbf{r})$ where $\delta_D(\mathbf{r})$ is a $D$-dimensional Dirac delta function, the physical dimension of $G_D$ can be obtained as $\left[G_D(\xi,\xi')\right] = \mathcal{L}^{2-D}$, where $\mathcal{L}$ denotes the fundamental dimension of length and $\left[X\right]$ denotes the unit of physical quantity $X$. Correspondingly, the partial derivative, $\partial G_D/\partial \xi$, in Eq. (16) can be non-dimensionalized as
\begin{equation}
\frac{\partial G_D(\xi, \xi')}{\partial \xi} = L^{2-D} \frac{\partial \mathcal{G}_D(\xi, \xi')}{\partial \xi} ,
\end{equation}
where $\mathcal{G}_D(\xi, \xi')$ is a dimensionless Green's function. We now rewrite Eqs. (15) and (16) as
\begin{subequations}
	\begin{gather}
	J_{D} f_0\int_0^1 \frac{d\xi'}{n_{\nu}(\xi')f_\nu} = e\mu_0\frac{V}{L} , \\
	1 = \left(\frac{ e^2 \mu_0 L^{2-D} }{ \epsilon f_0 J_D }  \right) n_{\nu}(\xi) f_\nu \int_0^1 d\xi' \frac{\partial \mathcal{G}_{D}(\xi,\xi')}{\partial \xi}n(\xi') , 
	\end{gather}
\end{subequations}
where $f_\nu\equiv f\left[n_\nu(\xi)\right]$ for simplicity. 
A direct inspection of Eq. (18a) shows that after the integral $\int d\xi' (\cdots)$ is fully converted into a dimensionless form, the $J_D$-$V$ and $J_D$-$L$ scaling relations can be unambiguously determined. In this case, $\int d\xi' (\cdots)$ becomes a \emph{dimensionless numeric factor} that does not play any roles in the $J_D$-$V$ and $J_D$-$L$ scaling relations.  

The non-dimensionalization of Eq. (18a) can be accomplished by appropriately regrouping the constant term in Eq. (18b), i.e. $\left( e^2\mu_0 L^{D-2}/ \epsilon f_0 J_D\right)$ into each of the $n_\nu(x)$ and $f_\nu$ terms in the right-hand side of Eq. (18b) such that dimensionless terms $\mathcal{N}_\nu(\xi)$ and $ \mathcal{F}_\nu$ can be defined, respectively, for $n_\nu(\xi)$ and $f_\nu$. 
In general, the regrouping of $\left( e^2\mu_0 L^{D-2}/\epsilon f_0 J_D\right)$ can be expressed in an arbitrary form of
\begin{equation}
\frac{ e^2 \mu_0 L^{2-D} }{ \epsilon f_0J_D } \equiv  \mathcal{A}_{J_D}(f_\nu) \tilde{\mathcal{A}}_{J_D}(f_\nu) \mathcal{B}_{L,D}(f_\nu) \tilde{\mathcal{B}}_{L,D}(f_\nu) 
\end{equation}
where $\mathcal{A}_{J_D}(f_\nu)$ and $\tilde{\mathcal{A}}_{J_D}(f_\nu)$ are terms containing $J_D$ and $\mathcal{B}_{L,D}(f_\nu) $ and  $\tilde{\mathcal{B}}_{L,D}(f_\nu) $ are terms containing $L^{2-D}$. 
The roles of $\mathcal{A}$'s and $\mathcal{B}$'s are to pair up with $n(\xi)$ and $f_\nu$ in Eq. (18b) such that the resulting terms are dimensionless. 

In the following, we suppress the argument of $\mathcal{A}$'s and $\mathcal{B}$'s for simplicity. As the explicit form of $f_\nu$ determines the regrouping of $\left( e^2\mu_0 L^{D-2}/\epsilon f_0 J_D\right)$, $\mathcal{A}$'s and $\mathcal{B}$'s are both $f_\nu$-dependent. Furthermore, $\mathcal{A}$'s are $D$-independent and $\mathcal{B}$'s are $D$-dependent as $L^{2-D}$ is deliberately distributed only into $\mathcal{B}$'s. 
We can now recast Eq. (18b) as 
\begin{equation}
1 = \mathcal{N}_\nu(\xi) \mathcal{F}_\nu \int_0^1 d\xi' \frac{\partial \mathcal{G}_D}{\partial \xi} \mathcal{N}(\xi') 
\end{equation}
where all terms are dimensionless via the following grouping
\begin{eqnarray}
\mathcal{N}_\nu(\xi) &\equiv& \left(\mathcal{A}_{J_D} \mathcal{B}_{L,D} \right) n_\nu(\xi), \nonumber \\
\mathcal{F}_\nu &\equiv& \left(\tilde{\mathcal{A}}_{J_D} \tilde{\mathcal{B}}_{L,D} \right)  f_\nu.
\end{eqnarray}
With $\mathcal{N}_\nu(\xi)$ and $\mathcal{F}_\nu$ now being dimensionless, Eq. (18a) can be rewritten as:
\begin{equation}
J_D  \mathcal{A}_{J_D} \tilde{\mathcal{A}}_{J_D}\mathcal{B}_{L,D} \tilde{\mathcal{B}}_{L,D} \int_0^1 \frac{d\xi'}{\mathcal{N}_{\nu}(\xi')\mathcal{F}_\nu}  = e\mu_0 \frac{V}{ L},
\end{equation} 
or more compactly as
\begin{equation}
J_D  \mathcal{A}_{J_D} \tilde{\mathcal{A}}_{J_D} = \psi_{\mathcal{G}_D} \frac{e \mu_0}{\mathcal{B}_{L,D}\tilde{\mathcal{B}}_{L,D}}\frac{V}{L},
\end{equation} 
where $\psi_{\mathcal{G}_D}$ is a dimensionless numeric factor dependent on the $D$ and $\mathcal{G}_D$, i.e.
\begin{equation}
\psi_{\mathcal{G}_D} \equiv  \left( \int_0^1 \frac{d\xi'}{\mathcal{N}_{\nu}(\xi')\mathcal{F}_\nu} \right)^{-1},
\end{equation}
which can be explicitly solved from Eq. (18b) and it affects only the overall magnitude of SCLC without affecting its voltage and length scaling relations. 
Thus the $J_D$-$V$ and $J_D$-$L$ scaling relations are determined by
\begin{equation}
J_D  \mathcal{A}_{J_D} \tilde{\mathcal{A}}_{J_D} \propto \frac{1}{\mathcal{B}_{L,D}\tilde{\mathcal{B}}_{L,D}}\frac{V}{L}.
\end{equation}

\subsection{Derivation of 2D thin-film SCLC scaling relations}

Equation (25) represents a \emph{universal SCLC scaling relations} for uniform SCLC injection into either a $D=1$ (bulk) or $D=2$ (thin film) based solid of length $L$ with arbitrary density-dependent mobility $\mu[n_\nu(\xi)]$. 
Several remarkable properties can be extracted from Eq. (25): 
(i) The $J_D$-$V$ scaling, i.e. $J_D  \mathcal{A}_{J_D} \tilde{\mathcal{A}}_{J_D} \propto V$, is determined solely by the $\mu[n_\nu(\xi)]$ and is \emph{completely independent on the device geometry ($\mathcal{G}_D$) and dimensionality ($D$)}; (ii) the $J_D$-$L$ scaling, on the other hand, is affected by both $f_\nu$ and $D$; (iii) For a fixed $D$ (= 1 for bulk or = 2 for thin film), the geometry of contacts affects only $\psi_{\mathcal{G}_D}$ and hence both SCL $J_D$-$V$ and $J_D$-$L$ scaling are universal independent on contact.

From Eq. (19), we see the constant term carriers a length scale dependence of $L^{2-D}$, and thus Eq. (19) will be independent of $L$ for $D=2$ (for a thin film setting) resulting in $\mathcal{B}_{L,2}=\tilde{\mathcal{B}}_{L,2} =1$. 
In this 2D thin film limit, Eq. (23) gives the SCLC relation for a thin film:
\begin{equation}
J_2 \mathcal{A}_{J_2}\tilde{\mathcal{A}}_{J_2} = \psi_{\mathcal{G}_2} e\mu_0 \left(\frac{V}{L}\right).
\end{equation}
Note that Eq. (26) includes that both $J_2$-$V$ and $J_2$-$L$ follow the same scaling relation. 

Togther with property (i) and Eq. (26), it shows a powerful tool that can be used to directly map the scaling relation of a simple bulk SCLC model into 2D thin-film SCLC model.
By virtual of Property (i), we conclude that the voltage scaling for thin-film ($J_2$-$V$) is identical to the bulk $J_1$-$V$ scaling, thus the voltage scaling obtained in Sec. II are valid for thin film as well.
For a thin film, Eq. (26) also dictates that the length ($J_2$-$L$) scaling relation is identical to the $J_2$-$V$.

To summarize this session, we provide a rigorous derivation of scaling laws (both voltage and length) for uniform SCL injection into a 2D thin film seeting. 
These properties allow the $J_2$-$V$ and $J_2$-$L$ scaling relations of a 2D thin-film to be fully determined from a simple 1D bulk SCLC model that have been shown in Sec. II.
The full $J_2$-$V$ and $J_2$-$L$ scaling relations are thus obtained without the need of explicitly solving the complicated coupled equations in Eq (18). 
In Appendix A, two examples of 2D thin-film SCLC are analyzed using our simplified formalism developed here. 

\subsection{Derivation of SCLC scaling relations and full numerical solutions of Eq. (18) for 2D Dirac semiconductor with traps}

The relativistic SCLC scaling relation of Dirac semiconductor in 2D thin-film geometry can be readily determined by using the simple derivation developed above. Since the 1D SCLC scaling relations takes the form of $J_1 \propto V^{3/2}$ and $J_1 \propto V^2$ respectively for the ultra-relativistic and non-relativistic regimes, for 2D thin film, our simple analysis yields
\begin{equation}
J_2 \propto \left(\frac{V}{L}\right)^\alpha ,
\end{equation}
where $\alpha$ = 2 and $\alpha$ = 3/2 are for the non-relativistic and ultra-relativistic limit, respectively. 
In the intermediate regime, the scaling follows an approximate power-law form with $\alpha$ varies continuously from $2\to3/2$ akin to Fig. 2.

\begin{figure}
	\includegraphics[scale=.9]{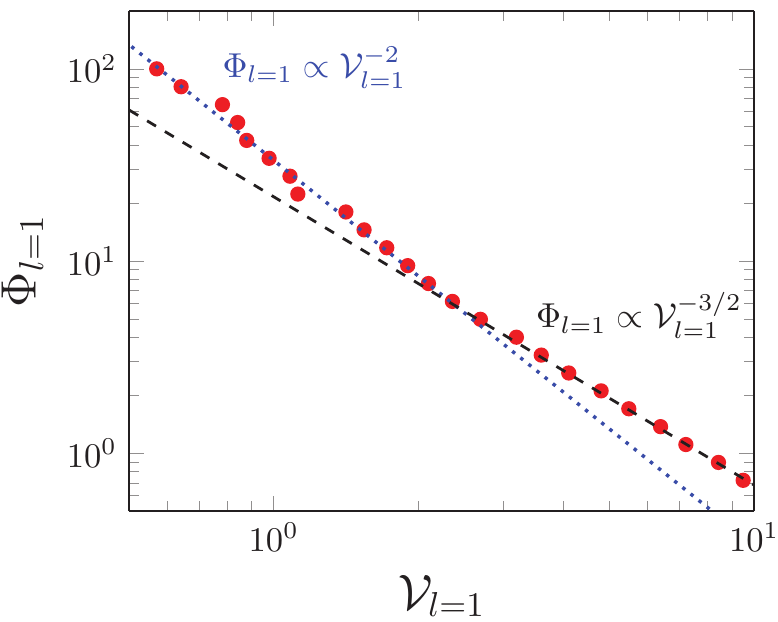}
	\caption{Numerical solution of the trap-free ($l=1$) 2D thin film of Dirac semiconductor. The dashed and dotted lines denotes $\Phi_{l=1} \propto \mathcal{V}_{l=1}^{-2}$ and $\Phi_{l=1} \propto \mathcal{V}_{l=1}^{-3/2}$, respectively. As $\Phi_{l=1} \propto 1/J_2$ and $\mathcal{V}_{l=1} \propto V/L$, the small-$\mathcal{V}_{l=1}$ and large-$\mathcal{V}_{l=1}$ regime corresponds to $J_2\propto (V/L)^2$ (non-relativistic) and $J_2 \propto (V/L)^{3/2}$ (ultra-relativistic), respectively. Note that the data points exhibits oscillations at small $\mathcal{V}_{l=1}$ due to numerical error.}
\end{figure}

To verify Eq. (27), the relativistic SCLC in 2D thin-film geometry with inclusion of exponential traps is explicitly solved (the full derivation is presented in Appendix B). In the presence of exponential traps, the relativistic SCLC in a 2D thin-film Dirac semiconductor is
	\begin{subequations} \label{2D}
		\begin{align}
		\frac{n_s^l(\xi)\left(1-\xi^2\right)^{1/2}}{\sqrt{C_ln_s^l(\xi) + n_c}} &= \int^1_{-1} \left( \frac{2\gamma e^2 C_l}{J_2\epsilon} \frac{n_s(\xi')\left(1-\xi'^2\right)^2}{\xi-\xi'} \right. \nonumber \\
			& \left.+\frac{\sqrt{C_ln_s^l(\xi)+n_c}}{\pi n_s^l(\xi')}\right)d\xi',  \\
		V &= \frac{J_2L}{\gamma eC_l} \int^1_{-1} \frac{\sqrt{C_ln_s^l(\xi) +n_c}}{n_s^l(\xi)} d\xi,
		\end{align}
	\end{subequations} 
where $l \equiv T_c/T >1$, $n_s(\xi)$ is the 2D carrier density, $n_c\equiv \rho_c/e$, $C_l\equiv N_0/N_t^l$ with $N_0$ as the effective density of states at the conduction band-edge and $N_t$ as the trap density, $\xi$ and $\xi'$ are dimensionless variables. 
In the ultra-relativistic and non-relativistic SCLC regimes, we obtain $J_2\propto \left(V/L\right)^{{l/2+ 1}}$ and $J_2\propto \left(V/L\right)^{l+1}$, respectively.
By setting $l=1$ (which corresponding to a trap-free case), we obtain $J_2 \propto (V/L)^2$ (non-relativistic) and $J_2 \propto (V/L)^{3/2}$ (ultra-relativistic), thus confirming the simple derivation in Eq. (27). This agreement demonstrates that the unconventional $J_D$-$V$ scaling of $3/2 < \alpha < 2$ is a universal signature of the relativistic charge carrier dynamics in both bulk and 2D thin-film geometries. 

In comparison to prior works in Ref. \cite{mos2}, the variation of $\alpha\approx1.7$ to $\alpha \approx 3$ with decreasing temperature was attributed to the transition from $T < T_c$ (valid) to $T > T_c$ (invalid). 
Our relativistic SCLC model with exponential traps presented here is however ever to take account for such temperature dependence without imposing the invalid $T_c<T$ condition.

To further confirm that the analytical relativistic SLCC scaling relation obtained above, the integral equation in Eq. (28a), which belongs to the class of \emph{nonlinear Cauchy singular integral equation} \cite{ladopoulos} is numerically solved for the trap-free case of $l=1$. 
The numerical solved $n_s(\xi)$ [from Eq. (28a)] is then integrated in Eq. (28b) to obtain the $J_2$ as a function of $V$. 
For simplicity, Eq. (28) is solved in terms of a dimensionless variables, i.e. $\mathcal{V}_{l=1} \propto V/L$ and $\Phi_{l=1} \propto 1/J_2$ [see Appendix C and Eqs. (C6) and (C7) for the definition of $\mathcal{V}_{l}$ and $\Phi_{l}$]. 
The numerical results (red circles) of $\Phi_{l=1} (\propto 1/J_2$) as a function of $\mathcal{V}_{l}$ is shown in Fig. 5 show good agreement with the derived scaling laws: $\Phi_{l=1} \propto \mathcal{V}_{l=1}^{-2}$ (dotted lines) at small voltage of $\mathcal{V}_{l=1} < 2$ and $\Phi_{l=1} \propto \mathcal{V}_{l=1}^{-3/2}$ at large voltage of of $\mathcal{V}_{l=1} > 2$. 
Thus, the comparison confirms the two corresponding analytical scaling laws [see Eq. (27)] for space charge limited conduction in a 2D thin-film Dirac solid with finite bangap: $J_2\propto (V/L)^2$ and $J_2 \propto (V/L)^{3/2}$ respectively for the non-relativistic and ultra-relativistic limits. 
More importantly, the unconventional relativistic SCLC scaling of $3/2<\alpha<2$ is unambiguously confirmed for the 2D thin-film Dirac semiconductor and is in agreement with experiments \cite{hbn,mos2}.

Finally, we discuss the screening effect on the relativistic SCLC in 2D Dirac materials. In 2D Dirac materials, the charge transport is sensitively influenced by the substrate screening and excess charge screening induced by gate-electrode in field-effect-transistor geometry. Despite these screening effects, SCLC was unambiguously observed in experiments as reported in Refs. [31] and [32]. These experimental observations suggest that the screening effect cannot entirely remove SCLC in 2D materials. In a previous theoretical work \cite{jena}, it is demonstrated that the surrounding dielectric screening will affect the transport properties of 2D thin film by modifying the relaxation time $\tau$ and a significant mobility enhancement can be achieved via high-κ dielectric substrate with vanishingly thin membrane. This theoretical prediction was experimentally confirmed in monolayer MoS2 with high-κ substrate\cite{radisaljevic}. In relevance to our SCLC model, we point out that as the substrate screening effect alters only the relaxation time $\tau$ which comes into the SCLC picture as a proportionality constant, it can be reasonably expected that only the magnitude of the SCLC will be altered while the new scaling laws reported here will remain unchanged. A microscopic theory of substrate screening can be formulated via first-principle calculation that takes into account the complex many-body interactions at the interface between 2D materials and the substrate\cite{kharche}. The complete microscopic quantum picture od dielectric screening is beyond the scope of this work.


\section{Conclusion} 

In summary, we have proposed a theory of relativistic space charge limited conduction (SCLC) in Dirac solids with new scaling laws for both bulk and thin file model.
For the one-dimensional (1D) bulk model, the scaling laws are $J_1 \propto V^{\alpha}/L^{\beta}$ with $3/2<\alpha<2$ and $2<\beta<3$. 
For 2D thin film model, we have $J_2 \propto (V/L)^\alpha$ for uniform SCLC injection with $\alpha$ remaining the same as the case of 1D bulk model under the assumption of density dependent mobility.
Both scaling laws have been verified with numerical calculations and have good agreements with experimental results.
The important finding from this paper is the new voltage scaling of $\alpha<2$ which is a signature of the massive Dirac fermions in the 2D Dirac materials, and cannot be explained by using the traditional SCLC models derived decades ago for traditional materials.
The inconsistencies in using such traditional SCLC models with unjustified traps condition to fit the experimental measurement is questionable.
Our results represent a new class of relativistic space-charge phenomena in Dirac solids, which may be used to model Dirac-based devices operating in space-charge-limited regime and may also be used as a tool to extract useful parameters by fitting the analytical equations with measurements.
The relativistic SCLC model should generate unconventional SCL photocurrent response \cite{SCLPC} in relativistic Dirac semiconductor such as MoS$_2$.
The widely-studied photoresponse of MoS$_2$ \cite{mos2_ph} can be readily used as an additional platform to verify the proposed relativistic SCLC model here.

\begin{acknowledgments} 
	
We thank Subhamoy Ghatak for providing us the experimental data of MoS$_2$ and insightful discussions.
We thank Chun Yun Kee, Kelvin J. A. Ooi and Shi-Jun Liang for helpful discussions. 
This work is supported by Singapore Ministry of Education T2 grant (T2MOE1401) and USA AFOAR AOARD Grant (FA2386-14-1-4020). 

\end{acknowledgments}


\begin{appendix}
	
	\section{Two examples of 2D thin-film SCLC using the simplified formalism}

	In this Appendix, we illustrates the simple derivation of the 2D SCLC scaling relations developed in Sections III-A and III-B using two examples.
	In the first example, we consider a trivial case with $f_\nu/f_0 = 1$, i.e. the mobility is independent of carrier density. For bulk model of  $D=1$, Eq. (18b) becomes
	\begin{equation}
	1 = \left(\frac{e^2\mu_0L}{\epsilon J_1}\right) \int_0^1 d\xi' \frac{\partial \mathcal{G}_1(\xi,\xi')}{d\xi}n(\xi'),
	\end{equation}
	which can be fully non-dimensionalized by defining $\mathcal{A}_J  = (e^2\mu_0/\epsilon J_1)^{1/2}$, $\tilde{\mathcal{A}}_{J_1} = 1$, $\mathcal{B}_{L,1} = L^{1/2}$ and $\tilde{\mathcal{B}}_{L,1} = 1$. From Eq. (23), we obtained $J_1 \left(e^2\mu_0/\epsilon J_1\right)^{1/2}= \psi_{\mathcal{G}_1} e\mu_0  V/L^{3/2}$, which can be rearranged to give the well-knwon bulk MG law of
	\begin{equation}
	J_1 = \psi_{\mathcal{G}_1}^2\epsilon\mu_0\frac{V^2}{L^3}.
	\end{equation}
	The numerical factor can be solved as $\psi_{\mathcal{G}_1}^2 = 9/8$ via Eq. (24) by using a 1D Green's function \cite{grinberg}. We can now map the $J_1$-$V$ bulk SCLC scaling relation to the 2D thin-film case, which yields $J_2 \propto V^2$. Furthermore, as $J_2$-$L$ scales equally with $J_2$-$V$, the 2D thin film SCLC scaling relation can now be fully determined as $J_2 \propto (V/L)^2$. One can verify this scaling relation by explicitly solving Eq. (26) with $\mathcal{A}_{J_2} = (e^2\mu_0/J_2)^{1/2}$ and $\tilde{\mathcal{A}}_{J_2} = 1$. This gives $J_2 \left(e^2\epsilon\mu_0/\epsilon J_2\right)^{1/2} = \psi_{\mathcal{G}_2} e\epsilon \mu_0 V/L$, which can be rearranged to give the well-known 2D thin-film SCLC \cite{grinberg}, i.e.
	\begin{equation}
	J_2 = \psi_{\mathcal{G}_2} \epsilon \mu_0 \left(\frac{V}{L}\right)^2,
	\end{equation}
	where $\psi_{\mathcal{G}_2}$ is a $\mathcal{G}_2$-dependent numeric factor. 
	
	In the second example, we consider a carrier density dependent mobility in a power-law form, i.e. $f_\nu = n_\nu(\xi)^{l-1}$ and $f_0 = n_0^{l-1}$ where $n_0$ and $l$ are some constants. This particular form of $\mu$ is equivalent to Mark-Helfrich's exponential-trap model with $l>1$. For this particular form of $f_\nu/f_0$, Eq. (18b) can be fully non-dimensionalized by re-grouping the constant factor $(e^2\epsilon\mu_0L/n_0^{l-1}J_1)$ via the following definitions: $\mathcal{A}_{J_1} = (e^2\epsilon\mu_0/n_0^{l-1}J_1)^{1/(l+1)} $, $\tilde{\mathcal{A}}_{J1} = (e^2\epsilon\mu_0/n_0^{l-1}J_1)^{(l-1)/(l+1)}$, $\mathcal{B}_{L,1} = L^{1/(l+1)}$ and $\tilde{\mathcal{B}}_{L,1} = L^{(l-1)/(l+1)}$. The bulk SCLC can then be obtained from Eq. (23) as
	\begin{equation}
	J_1 \left( \frac{e^2\mu_0}{n_0^{l-1}\epsilon J_1} \right)^{\frac{l}{l+1}} = \psi_{\mathcal{G}_1} e\mu_0 \frac{V}{L^{\frac{2l+1}{l+1}}},
	\end{equation}
	which can be simplified as 
	\begin{equation}
	J_1 = \psi_{\mathcal{G}_1}^{l+1} \left(en_0^l\right)^{l-1}\epsilon^l \mu_0 \frac{V^{l+1}}{L^{2l+1}},
	\end{equation}
	and is in agreement with the Mark-Helfrich's exponential trap model \cite{mark}. To generalize the bulk SCLC to the case of 2D thin film, we again utilize the facts that: (i) $J_2$-$V$ follows the same scaling as $J_1$-$V$; and (ii) $J_2$-$L$ scales equally with $J_2$-$V$  This gives $J_{2} \propto (V/L)^{l+1}$, which is in agreement with the explicitly solution of Eq. (26), i.e.
	\begin{equation}
	J_2 = \psi_{\mathcal{G}_{2} }^{l+1} \left(en_0^l\right)^{l-1} \epsilon^l \mu_0 \left(\frac{V}{L}\right)^{l+1}.
	\end{equation}

	\section{Derivation of 2D relativistic SCLC model}
	
	In this Appendix, we provide a full derivation of the Mark-Helfrich's SCLC model and Dirac semiconductor in 2D thin-film geometry based on Grinburg's formalism \cite{grinberg}.

	\subsection{Mark-Helfrich's trap model of SCLC in 2D thin-film geometry}
	
	In the presence of traps that follows an exponential energy distribution \cite{mark}, the free and trapped carrier densities are related by  $n_f(x) = \left(N_0/N_t^l\right) n_s^l(x) \equiv C_l n_s^l(x)$ where $n_f(x)$ is the free carrier density, $n_s(x)$ is the trapped carrier density, $N_0$ is the effective density of states  at the conduction band edge, $N_t$ is the trap density and $C_l \equiv N_0/N_t^l$. Here, $l \equiv T_c/T \geq 1$ where $T_c$ is a characteristic temperature representing the exponential spread in energy of the traps. The charge density in a 2D thin-film is given as
	\begin{equation}
	\rho(x,y) = e\delta(y)\left[-n_s(x) + P_s\delta(L-x)\right] ,
	\end{equation}
	where $\delta(y)$ is a Dirac delta function and $P_s = \int_0^L n_s(x) dx$ is the charge density induced on the annode by the total $n_s(x)$ residing in the thin film. Note that $y$ represent the direction that is out-of-plane of the thin film. For thin-edge contacts, the corresponding Green's function is 
	\begin{equation}
	G(x-x',y-y') = -\frac{1}{2\pi} \ln\left[ (x-x')^2 + (y-y')^2 \right]^{1/2},
	\end{equation}
	and the scalar potential can then be solved as
	\begin{widetext}
		\begin{equation}
		\phi(x,y) = -\frac{1}{2\pi} \int^\infty_{-\infty} dy' \int^L_0 dx' \ln\left[ (x-x')^2 + (y-y')^2 \right]^{1/2} \left(-\frac{4\pi e}{\epsilon}\right) \delta(y') \left[ -n_s(x) + P_s\delta(L-x) \right].
		\end{equation}
		Simplify $\phi(x,y=0)$ and knowing that $E_x(x,0) = -d\phi/dx$, we obtain
		\begin{equation}
		E_x(x,0) = \frac{2e}{\epsilon(L-x)} \int^L_0 \frac{L-x'}{x-x'} n_s(x') dx'.
		\end{equation}
		By defining $\xi = x/L$ and $\xi' = x'/L$, we obtained
		\begin{equation}
		E_x(\xi,0) = \frac{2e}{\epsilon(1-\xi)} \int^1_0 \frac{1-\xi'}{\xi-\xi'} n_s(\xi') d\xi'.
		\end{equation}
		We now consider a current density equation in Drude's form, i.e.
		\begin{equation}
		J = en_f(x) \mu E_x(x,0).
		\end{equation}
		By combining Eq. (B5) and (B6), we obtain
		\begin{equation}
		1 = \frac{2e^2 \mu C_l}{J \epsilon} \frac{n_s(x)^l}{1-\xi} \int^1_0 \frac{1-\xi'}{\xi-\xi'} n_s(\xi') d\xi' .
		\end{equation}
		Equation (B7) can be rearranged as followed:
		\begin{equation}
		1 = \left(\frac{2e^2\mu C_l}{J\epsilon}\right)^{\frac{l}{l+1}} \frac{n_s^l(\xi)}{1-\xi} \int^1_0 \frac{1-\xi'}{\xi-\xi'} \left(\frac{2e^2\mu C_l}{J\epsilon}\right)^{\frac{1}{l+1}} n_s(\xi') d\xi' .
		\end{equation}
	\end{widetext}
	By defining
	\begin{equation}
	\nu_s(\xi) \equiv \left(\frac{2e^2\mu C_l}{J\epsilon}\right)^{\frac{1}{l+1}} n_s(\xi) ,
	\end{equation}
	Eq. (B8) becomes
	\begin{equation}
	1 = \frac{\nu_s^l(\xi)}{1-\xi} \int^1_0 \frac{1-\xi'}{\xi-\xi'}  \nu_s(\xi') d\xi' ,
	\end{equation}
	which is an integral equation that can be solved to obtain $\nu(\xi)$. The bias voltage can be obtained from $
	V = \int^1_0 E_x(\xi,0) d\xi$ and Eq. (B6) as
	\begin{equation}
	V = \frac{JL}{e\mu C_l} \int_0^1 \frac{d\xi}{n^l_s(\xi)}.
	\end{equation}
	To obtain the exponential trap-limited SCLC in 2D thin film geometry with edge-contact, Eqs. (B9) and (B11) are combined to give
	\begin{equation}
	V = \frac{JL}{\epsilon \mu C_l}  \left(\frac{2e^2\mu C_l}{J\epsilon}\right)^{\frac{1}{l+1}} \int_0^1 \frac{d\xi}{\nu_s(\xi)}.
	\end{equation}
	With the definition of $\lambda \equiv \int_0^1d\xi/\nu_s(\xi)$, which is a constant that can be solved from the integral equation in Eq. (B10), we obtain
	\begin{equation}
	J = \left(\frac{\epsilon}{2\lambda}\right)^l e^2 \mu C_l \left(\frac{V}{L}\right)^{l+1}.
	\end{equation}
	\begin{widetext}	
		Equation (B13) gives the exponential trap-limited SCLC of a 2D thin film with edge-contact geometry [see Fig. 4(b)]. For strip-geometry [see Fig. 4(c)], the electric field is given as
		\begin{equation}
		E_x(\xi) = \frac{2}{(1-\xi^2)^{1/2}} \left( \frac{e}{\epsilon} \int^1_{-1} \frac{n_s(\xi')\left(1-\xi'^2\right)^{1/2}}{\xi-\xi'}d\xi' +\frac{V}{\pi L} \right),
		\end{equation}
		where $\xi \equiv (2x-L)/L$ and $\xi' \equiv (2x'-L)/L$. Using similar procedure, we obtain
		\begin{equation}
		\frac{J}{e\mu} = \frac{2C_ln_s^l(\xi)}{(1-\xi^2)^{1/2}} \left( \frac{e}{\epsilon} \int^1_{-1} \frac{n_s(\xi')\left(1-\xi'^2\right)^{1/2}}{\xi-\xi'}d\xi' +  \frac{1}{\pi L} \frac{JL}{e\mu C_l} \int^1_0\frac{d\xi}{n^l_s(\xi)} \right),
		\end{equation}
		which can be simplified to 
		\begin{equation}
		1 = \frac{\nu_s^l(\xi)}{(1-\xi^2)^{1/2}} \left(  \int^1_{-1} \frac{\nu_s(\xi')\left(1-\xi'^2\right)^{1/2}}{\xi-\xi'}d\xi' + \frac{1}{\pi} \int^1_{-1}\frac{d\xi}{\nu^l_s(\xi)} \right).
		\end{equation}
	\end{widetext}
	From Eq. (B12), the SCLC current density equation is obtained as
	\begin{equation}
	J = \left(\frac{\epsilon}{2\lambda'}\right)^l e^2 \mu C_l \left(\frac{V}{L}\right)^{l+1},
	\end{equation}
	where the numeric factor $\lambda' \equiv \int_{-1}^{1} d\xi \nu_s(\xi)$ can be obtained by solving $\nu_s(\xi)$ from Eq. (B16). In summary, the 2D thin-film uniform injection of SCLC in the presence of exponential traps follow the following scaling relation of
	\begin{equation}
	J \propto \left(\frac{V}{L}\right)^{l+1},
	\end{equation}
	for both edge-and strip-contact geometries. More importantly, this scaling relation is in agreement with Eq. (A6) obtained using the simplified formalism.

	\subsection{Relativistic SCLC model for 2D massive Dirac fermions}
	
	For 2D Dirac semiconductor, we obtain
	\begin{widetext}
		\begin{equation}
		1 = \frac{2\gamma e^2 C_l n_s^l(\xi) }{J \epsilon \sqrt{C_l n_s^l(\xi) + n_c}} \int^1_0 \frac{1-\xi'}{\xi-\xi'} n_s(\xi')d\xi',
		\end{equation}
		and
		\begin{equation}
		\frac{\sqrt{C_ln_s^l(\tilde{\xi}) + n_c}\left(1-\tilde{\xi}^2\right)^{1/2}}{n_s^l(\tilde{\xi})} = \int^1_{-1} {\left( \frac{2\gamma e^2 V_l}{J\epsilon} \frac{n_s(\tilde{\xi}')\left(1-\tilde{\xi}'^2\right)^2}{\tilde{\xi}-\tilde{\xi}'} + \frac{\sqrt{C_ln_s^l(\tilde{\xi})+n_c}}{\pi n_s^l(\tilde{\xi}')}\right)d\tilde{\xi}'},
		\end{equation}
	\end{widetext}
	respectively for edge-contact and strip-contact. The applied bias voltage for edge-and strip-contact geometries become, respectively,
	\begin{equation}
	V = \frac{JL}{\gamma eC_l} \int^1_{0} \frac{\sqrt{C_ln_s^l(\xi) +n_c}}{n_s^l(\xi)} d\xi,
	\end{equation}
	and
	\begin{equation}
	V = \frac{JL}{\gamma eC_l} \int^1_{-1} \frac{\sqrt{C_ln_s^l(\tilde{\xi}) +n_c}}{n_s^l(\tilde{\xi})} d\tilde{\xi}.
	\end{equation}
	The coupled Eqs. (B19) to (B22) can be solved to obtain the relativistic SCLC in 2D thin-film geometry. Equations (B19) to (B22) has to be solved numerically. Nonetheless, in the non-relativistic and ultra-relativistic limits, semi-analytical scaling relations can be derived. We first consider the non-relativistic limit of $n_c\gg n_s^l(\xi)$ for all $\xi$ with edge-contacts, Eqs. (B19) and (B21) can be approximated, respectively, by
	\begin{equation}
	1 = \frac{2\gamma e^2 C_l n_s^l(\xi)}{J \epsilon n_c^{1/2}} \int^1_0 \frac{1-\xi'}{\xi - \xi'} n_s(\xi') d\xi',
	\end{equation}
	and
	\begin{equation}
	V = \frac{JLn_c^{1/2}}{\gamma eC_l} \int^1_0 \frac{d\xi}{n_s^l(\xi)},
	\end{equation}
	By defining
	\begin{equation}
	\nu_s(\xi) \equiv \left(\frac{2\gamma e^2 C_l}{J \epsilon n_c^{1/2}}\right)^{\frac{1}{l+1}} n_s(\xi),
	\end{equation}
	we obtain
	\begin{equation}
	J = \frac{1}{\lambda^{l+1}} \left(\frac{\epsilon}{2}\right)^l\frac{\gamma  C_l  }{ n_c^{l/2}  e^{2l-1}} \left(\frac{V}{L}\right)^{l+1},
	\end{equation}
	where $\lambda \equiv \int^1_0 d\xi/\nu_s^l(\xi)$ is a numerical factor which can be solved from the nonlinear integral equation in Eq. (B19). By setting $l=1$, the current-voltage scaling relation agrees with the 1D bulk model as shown in Eq. (5a) of the main text. The current-voltage scales equally with the current-length which is also in agreement with the simplified derivation of 2D thin-film SCLC scaling relation presented in Eq. (27) of the main text.
	
	In the ultra-relativistic limit of $n_c\to 0$, Eqs. (B19) and (B21) become, respectively,
	\begin{equation}
	1 = \nu_s^{l/2}(\xi) \int^1_0 \frac{1-\xi'}{\xi-\xi'} \nu_s(\xi')d\xi',
	\end{equation}
	and
	\begin{equation}
	V = \frac{JL}{\gamma e}\left(\frac{2\gamma e^2 C_l^{1/2}}{J\epsilon}\right)^{\frac{l/2}{l/2+1}} \int^1_0 \frac{d\xi}{\nu_s^{l/2}(\xi)},
	\end{equation}
	where 
	\begin{equation}
	\nu_s(\xi) \equiv \left(\frac{2\gamma e^2 C_l^{1/2}}{J\epsilon}\right)^{\frac{1}{l/2+1}} n_s(\xi),
	\end{equation}
	which can be rearranged to give
	\begin{equation}
	J = \frac{1}{\lambda'^{l/2 + 1}}\left(\frac{\epsilon}{2}\right)^{l/2} \frac{\gamma e^{1-l/2} }{C_l^{l/4}} \left(\frac{V}{L}\right)^{\frac{l}{2} + 1}.
	\end{equation}
	The numerical factor, $\lambda' \equiv \int^1_0 d\xi / \nu_s^{l/2} (\xi)$, can again be solved from Eq. (B19). For $l=1$, the current-voltage scaling relation agrees with the ultra-relativistic results in Eq. (5b) of the main text. In the intermediate regime,  the scaling relation can be approximated by
	\begin{equation}
	J \propto (V/L)^\Lambda,
	\end{equation}
	where $\Lambda = l/2+1$ and $\Lambda = l+1$.

	\begin{widetext}
		
		\section{Equations (B28) to (B32) in dimensionless form}	
		
		Equations. (B28) to (B32)	can be transformed into a dimensionless form for numerical solution in Fig. 5 of main text. For \emph{edge-contacts}, we obtain
		\begin{equation}
		1 = \frac{\Phi_l}{1-\xi} \frac{f_s^l(\xi)}{\sqrt{f_s^l(\xi) + 1}} \int^1_0 \frac{1-\xi'}{\xi-\xi'} f_s(\xi') d\xi',
		\end{equation}
		and
		\begin{equation} 
		\mathcal{V}_l = \frac{1}{\Phi_l} \int^1_0 \frac{\sqrt{f_s^l(\xi) + 1}}{f_s^l(\xi)} d\xi.
		\end{equation}
		
		For \emph{strip-contacts}, the dimensionless form give
		\begin{equation}
		1 = \frac{f_s^l(\xi)}{\sqrt{f_s^l(\xi) + 1}} \frac{1}{(1-\xi^2)^{1/2}} \int_{-1}^{1} \left( \Phi_l f_s(\xi') \frac{(1-\xi'^2)^{1/2}}{\xi-\xi'} + \frac{1}{\pi } \frac{\sqrt{f_s^l(\xi') + 1} }{f_s^l(\xi')} \right) d\xi',
		\end{equation}
		and
		\begin{equation}
		\mathcal{V}_l = \frac{1}{\Phi_l} \int^1_{-1} \frac{\sqrt{f_s^l(\xi)+1}}{f_s^l(\xi)} d\xi.
		\end{equation}	
		The dimensionless parameters are defined as
		\begin{equation}
		f_s \equiv \frac{C_l}{n_c}^{1/l} n_s(\xi),
		\end{equation}
		\begin{equation}
		\Phi_l \equiv \frac{2\gamma e^2}{J\epsilon} \left( \frac{n_c}{C_l} \right)^{1/l} \sqrt{n_c},
		\end{equation}
		and
		\begin{equation}
		\mathcal{V}_l \equiv \frac{\epsilon V }{2eL} \left( \frac{C_l}{n_c} \right)^{1/l},
		\end{equation}
		where $\Phi_l$ is current-and material-dependent parameter. 
	\end{widetext}

\end{appendix}


\begin{thebibliography}{99}
\bibitem{fn2} At the transitional regime where $I$ is approximately proportional to $V^{3/2}$, $V^3/I^2$ becomes constant. This leads to the occurrence of a peak in the $V^3/I^2$-$1/V$ plot.
\end{thebibliography}
\end{document}